\documentclass[prx,aps,twocolumn,superscriptaddress]{revtex4-2}

\usepackage[pdftex]{graphicx}
\usepackage{amsbsy,amssymb,amsmath,bm,mathtools}

\usepackage{xspace}
\usepackage{bm}
\usepackage{color}
 
\usepackage[mathscr]{euscript}

\usepackage{url}
\usepackage[section]{placeins}
\usepackage[colorlinks=true,linkcolor=blue,citecolor=blue]{hyperref}

\newcommand\norm[1]{\left\lVert#1\right\rVert}

\begin{document}


\title{Echo State network for coarsening dynamics of charge density waves}

\author{Clement Dinh}
\affiliation{Department of Physics, University of Virginia, Charlottesville, VA 22904, USA}

\author{Yunhao Fan}
\affiliation{Department of Physics, University of Virginia, Charlottesville, VA 22904, USA}

\author{Gia-Wei Chern}
\affiliation{Department of Physics, University of Virginia, Charlottesville, VA 22904, USA}

\date{\today}

\begin{abstract}
An echo state network (ESN) is a type of reservoir computer that uses a recurrent neural network with a sparsely connected hidden layer. Compared with other recurrent neural networks, one great advantage of ESN is the simplicity of its training process. Yet, despite the seemingly restricted learnable parameters, ESN has been shown to successfully capture the spatial-temporal dynamics of complex patterns. Here we build an ESN to model the coarsening dynamics of charge-density waves (CDW) in a semi-classical Holstein model, which exhibits a checkerboard electron density modulation at half-filling stabilized by a commensurate lattice distortion. The inputs to the ESN are local CDW order-parameters in a finite neighborhood centered around a given site, while the output is the predicted CDW order of the center site at the next time step. Special care is taken in the design of couplings between hidden layer and input nodes to ensure lattice symmetries are properly incorporated into the ESN model. Since the model predictions depend only on CDW configurations of a finite domain, the ESN is scalable and transferrable in the sense that a model trained on dataset from a small system can be directly applied to dynamical simulations on larger lattices. Our work opens a new avenue for efficient dynamical modeling of pattern formations in functional electron materials.  
\end{abstract}

\maketitle

\section{Introduction} 

\label{sec:intro}

Recurrent Neural Networks (RNNs) have become increasingly important tools in physics, particularly for modeling and analyzing systems that involve time-series data or dynamic processes~\cite{sherstinsky20,lipton15,Funahashi93,Mandic01}. The characteristic feature of RNNs that distinguishes them from the more widely used feedforward neural networks is that the connection topology possesses cycles. The presence of cycles means that an RNN may develop self-sustained temporal activation dynamics along its recurrent connection pathways, even in the absence of input. The quasi-self-sustained dynamical loops of a RNN can also serve as nonlinear filters that transform an input time series into an output time series. 


One of the main challenges in developing RNN models is the training complexity. This is because the presence of loops also makes the conventional gradient-descent-based optimizations unstable~\cite{doya92}.  A new paradigmatic approach to RNNs is based on the concept of reservoir computing (RC)~\cite{lukosevicius09,Tanaka19,Nakajima20}, which emphasizes fixed, dynamic reservoirs for processing temporal data. Two representative RC systems are the echo state network (ESN)~\cite{jaeger01} and the liquid state machine~\cite{maass02}. The reservoir in the former is based on continuous-time random neural networks, while the latter is biologically inspired and is based on discrete stochastic spiking neurons. Importantly, since the reservoir network remains fixed after initialization, the training complexity and computational cost of RC-based models are significantly reduced.


The architecture of ESN is especially well-suited for learning complex, yet deterministic, dynamics or time sequences~\cite{jaeger01,jaeger04,manjunath13}. An ESN consists of three main components: the input, output, and reservoir layers. The input layer is fully connected to the reservoir layer, which is a large randomly connected recurrent network. The reservoir neurons are usually activated using a non-linear activation function such as a sigmoid or ReLU. The activation function allows the reservoir to produce complex and non-linear responses to inputs. Finally, the output layer is a linear layer that takes the state of the reservoir and generates the final prediction or output of the network. As opposed to the reservoir, the output layer is trained. This means that the weights from the reservoir neurons to the output neurons are learned using a supervised learning algorithm (e.g., least squares or ridge regression), while the reservoir weights remain fixed.

The statistical nature of machine learning models means that dynamical behaviors predicted by RNNs will eventually diverge from actual trajectories because of prediction errors. This is a particularly important issue for modeling chaotic systems which are known to be very sensitive to small differences in initial conditions. Yet, it has been shown that a properly trained ESN can capture the long-term statistical patterns, or the so-called ``climate", of chaotic dynamics~\cite{Pathak18,Pathak17,Haluszczynski19,Roy22,Chattopadhyay20,Chen20}. These include the power spectrum of the time series, the correlation dimension, and the Lyapunov exponent. These studies highlight a tantalizing model-free approach for predicting complex or chaotic dynamics from experimental data. 

In this paper, we present a scalable machine learning (ML) framework based on ESN to achieve a model-free prediction of adiabatic phase ordering dynamics in complex electron systems. We demonstrate our approach to the coarsening dynamics of charge-density wave (CDW) orders in the square-lattice Holstein model~\cite{noack91,zhang19,chen19,hohenadler19}. Although the adiabatic evolution of a CDW state in the Holstein model can be numerically exactly solved in the semiclassical approximation, the calculation of forces that drive the CDW evolution requires solving an electron structure problem. Since this has to be done at every time step of the dynamical simulation, large-scale simulations of the CDW dynamics based on exact solutions are very time-consuming. We show that the ESN offers an efficient linear-scaling approach to this computationally difficult problem. 

Fundamentally, the linear scalability of our approach relies on the locality principle~\cite{kohn96,prodan05}, which in our case means that the update to a local CDW order parameter only depends on its immediate surroundings. Consequently, the ESN is designed to predict the time evolution of a local on-site CDW order parameter based on input from CDW configuration in a finite neighborhood at the previous time-step. A new configuration of the whole system is obtained by repeatedly applying the same trained ESN to every sites one the lattice. As a result, our ML model is transferable and scalable since a successfully trained ESN model can be applied to arbitrarily large system without rebuilding or retraining.

As in previous studies, the trained ESN can only generate accurate short-term results, while the long-term predictions inevitably deviate from the exact simulations. Nonetheless, the overall statistical behaviors such as the domain-growth law and dynamical scaling are accurately captured by the ESN models, similar to the correct modeling of the climate of chaotic systems discussed above. We demonstrate this benchmark for the phase ordering of the well-studied Time-Dependent Ginzburg-Landau (TDGL) equation~\cite{Bray1994,Onuki2002,Puri2009}. Applying the ESN approach to enable large-scale dynamics simulations, we find that the phase-ordering dynamics of the checkerboard CDW order, which is also described by an Ising order parameter on symmetry grounds, exhibit statistical behaviors that are distinctively different from the Allen-Cahn growth law expected for a non-conserved Ising order.

The rest of the paper is organized as follows. In Sec.~\ref{sec:review} after briefly reviewing the architecture of ESN, we present a scalable ML framework based on a local ESN model for phase ordering dynamics of broken symmetry phases. We apply it to study the phase-ordering dynamics of a non-conserved Ising order as described by the TDGL in Sec.~\ref{sec:TDGL}. We demonstrate that while ESN cannot produce exact long-term behaviors, it nonetheless captures the statistical coarsening behavior and correctly reproduces the expected Allen-Cahn domain growth law. In Sec.~\ref{sec:esn-cdw} we apply the ESN model to the adiabatic dynamics of CDW order in the semiclassical Holstein model. We discuss the exact diagonalization and the Langevin method for generating training dataset, and present the details and benchmark of the ESN model for the dynamics of CDW order.   Finally, we present a conclusion and outlook in Sec.~\ref{sec:conclusion}.


\section{ESN for phase ordering dynamics}

\label{sec:review}

\subsection{Echo state network: brief review}

\begin{figure*}
\centering
\includegraphics[width=1.99\columnwidth]{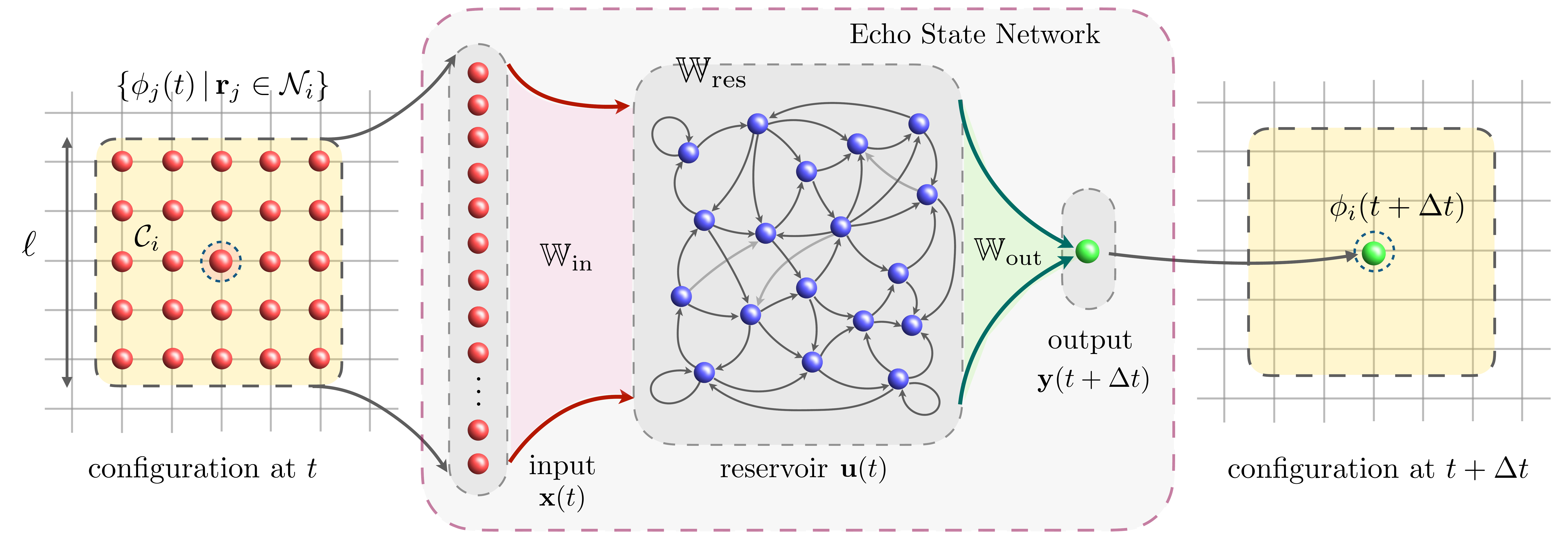}
\caption{Schematic diagram of a scalable ML framework based on echo state network (ESN) for coarsening dynamics a scalar order-parameter field $\phi(\mathbf r, t)$. The middle block shows the architecture of an ESN. The three main components are the input layer $\mathbf x(t)$, the reservoir neural net $\mathbf u(t)$, and the output neurons $\mathbf y(t)$. The reservoir is fully connected to both the input and output layers with weight matrices $\mathbb{W}_{\rm in}$ and $\mathbb{W}_{\rm out}$, respectively. The ESN is designed to predict the order parameter $\phi(\mathbf r_i, t+\Delta t)$ of a given lattice site-$i$ at the next time-step, which corresponds to the signal neuron at the output layer. The ESN prediction is based on the current local order-parameters $\phi(\mathbf r_j, t)$ within a finite neighborhood $\mathcal{N}_i$ of site-$i$. The neighborhood configuration is flattened into an array $\mathbf x(t)$ in the input layer of the ESN. }
    \label{fig:schematic}
\end{figure*}

The ESN architecture features three primary components: an input layer, an output layer, and a ``reservoir'' connected between them. Contrary to typical feedforward networks where inputs are fed sequentially through a series of hidden layers, the neurons in a reservoir are not arranged in a layered structure. Instead, the reservoir is represented by a random, fixed, high-dimensional recurrent neural network, which receives signals from all nodes in the input layer. Linear combinations of signals from all reservoir neurons are then assigned to the output layer. A schematic diagram of ESN is shown in the middle block of FIG.~\ref{fig:schematic}. Neurons in the reservoir layer can excite each other depending on the connections. In particular, the presence of multiple loops inside means that the reservoir can stay in a self-sustainable dynamical state even without driving signals from the input. Also importantly, the weights of the input and reservoir nodes in an ESN are randomly initialized and stay fixed throughout the entire training process while only the weights of the output nodes are adjusted which cuts down the training cost of this type of network significantly.  

The size of the reservoir is typically much larger than the size of the input and output layers to ensure that it has enough complexity to model the dynamics of the system. The recurrent connections give the reservoir a rich dynamical behavior. Specifically, a key requirement for the reservoir layer of an ESN is the echo state property, which means that the state of the reservoir must eventually ``forget'' its initial condition and rely solely on the recent inputs. In other words, the system's dynamics are primarily influenced by the inputs over time, rather than being dominated by the initial state. 


To describe the dynamical evolution of an ESN, we introduce an array or vector $\mathbf u(t) = [ u_1(t), u_2(t), \cdots, u_{N_{\rm res}}(t)]$ to denote signals of the reservoir neurons at time $t$, where $N_{\rm res}$ is the size of the reservoir network. Similarly, vectors $\mathbf x(t)$ and $\mathbf y(t)$ are used to represent signals of the input and output neurons, respectively. The ESN prediction is done in two stages. First, the reservoir neurons and the input neurons are nonlinearly transformed to get the reservoir state at the next time-step:
\begin{eqnarray} 
	\label{eq:dyn1}
	\textbf{u}(t+\Delta t) = f_{\rm av}\left( \mathbb{W}_{\rm res} \cdot \mathbf u(t) 
	+ \mathbb{W}_{\rm in} \cdot \mathbf x(t) + \mathbf b \right) + \bm \xi(t). \nonumber \\
\end{eqnarray}
Here $\mathbb{W}_{\rm res}$ is a $N_{\rm res}\times N_{\rm res}$ fixed weight matrix for the reservoir neurons, $\mathbb{W}_{\rm in}$ is a $N_{\rm res} \times N_{\rm in}$ matrix describing the weighted coupling of reservoir to the input neurons, $N_{\rm in}$ is the size of the input layer, $\mathbf b$ is a fixed bias vector, and $f_{\rm av}(x)$ is the activation function, often chosen to be $\tanh(x)$. Here we have also included random noises~$\bm \xi(t)$ to the excitation source of reservoir neurons, similar to the Langevin noise in molecular dynamics. Typically, this term is sampled from a uniform or a Gaussian distribution with a standard deviation of the order of $10^{-4}-10^{-2}$. It has been shown that the inclusion of such noise terms could increase the stability of the reservoir network during the training process to be discussed below~\cite{jaeger01}. For simplicity, the bias vector is set to zero in our implementation of the ESN. 

The next step is to map the reservoir states to the output layer via a simple linear transformation
\begin{eqnarray}
	\label{eq:dyn2}
	\mathbf y(t+\Delta t) =  \mathbb{W}_{\rm out} \cdot \mathbf u(t + \Delta t) ,
\end{eqnarray}
where $\mathbb{W}_{\rm out}$ is a $N_{\rm out}\times N_{\rm res}$ matrix describing the weighted connections between neurons of the two layers, and $N_{\rm out}$ is the number of output neurons.

As discussed above, both weight matrices $\mathbb{W}_{\rm res}$ and $\mathbb{W}_{\rm in}$ are randomly initialized and remain fixed throughout the training. The elements of the weight matrix $\mathbb{W}_{\rm out}$ are the only trainable parameters of an ESN. The training procedure begins by arbitrarily initializing the reservoir nodes and then driving the network using input data time sequences $\mathbf x_{\rm data}(t)$.  For a given training time series, it is necessary to ``wash out" the initial transient dynamics of the arbitrarily initialized reservoir. This is done typically by ignoring the output signal of the ESN until some transient time $T_0$ along with an appropriate scaling of the weights of $\mathbb{W}_{\rm res}$ (smaller weights increase damping). 
As the goal is to accurately predict the output data sequence $\mathbf y_{\rm data}(t)$, the training is equivalent to the minimization of a loss function
\begin{eqnarray}
	L = \sum_{t_k > T_0} \norm{ \mathbb{W}_{\rm out} \cdot \mathbf u(t_k) - \mathbf y_{\rm data}(t_k) }^2 + \epsilon \norm{ \mathbb{W}_{\rm out} }^2
\end{eqnarray}
where $\epsilon$ denotes a regularization constant, which prevents overfitting by penalizing large values of the fitting parameter. The minimization of $L$ with respect to $\mathbb{W}_{\rm out}$ is then translated to a problem of Ridge regression. Specifically, we define a $\mathbb{M}$ matrix of dimension $N_t \times N_{\rm res}$, where $N_t$ is the size of the time sequence in the above summation such that each column of $\mathbb{M}$ corresponds to the reservoir state $\mathbf u(t_k)$ at a particular time-step $t_k$, where $k = 1, 2, \cdots, N_t$. Similarly, a teaching matrix $\mathbb{R}$ of size $N_t \times N_{\rm out}$ is introduced such that each column of $\mathbb{R}$ corresponds to an output neuron data $\mathbf y_{\rm data}(t_k)$ at a given time $t_k$. The explicit solution of the optimal output weight matrix is given by
\begin{eqnarray}
	\label{eq:solve_Wout}
	\mathbb{W}_{\rm out} = \left( \mathbb{M}^{\rm t} \cdot \mathbb{M} + \epsilon \mathbb{I} \right)^{-1} \cdot \mathbb{M}^{\rm t} \cdot \mathbb{R}. 
\end{eqnarray}
It is noted that $\mathbb{W}_{\rm in}$ and $\mathbb{W}_{\rm res}$ should be properly scaled depending on the task while still being randomly initialized. Specifically, the absolute value of $\mathbb{W}_{\rm in}$ weights should reflect how much influence the input signal has on the network. Additionally, the reservoir matrix $\mathbb{W}_{\rm res}$  should be sparse and satisfy the echo state property, which requires its spectral radius $\rho$ to be less than 1. Here the spectral radius of a random weight matrix is defined as its largest absolute eigenvalue: $\rho(\mathbb{W}_{\rm res}) = \max\{|\lambda_1|, |\lambda_2|, \cdots, |\lambda_N| \}$, where $\lambda_i$ are eigenvalues of the random matrix. Practically, the initialization of a random weight matrix with a specific spectral radius $\rho < 1$ can be achieved by first generating a random weight matrix $\mathbb{W}^{(0)}_{\rm res}$ whose spectral radius is $\rho_0$, then rescaling it to produce the desired matrix $\mathbb{W}_{\rm res} = (\rho/ \rho_0) \mathbb{W}^{(0)}_{\rm res}$.  The spectral radius $\rho$ determines how well the model can adapt to its teacher dynamics. Namely, ESNs with $\mathbb{W}_{\rm res}$ characterized by a small spectral radius demonstrate better performance on learning temporally fast dynamics, while models with a large $\rho$ perform better on learning temporally slow dynamics.

\subsection{ML framework for phase ordering dynamics }

Here we present a scalable ML framework based on ESN for modeling coarsening dynamics of order-parameter field. While our proposed framework can be applied to general symmetry-breaking phases characterized by multiple order parameters, for convenience of presentation we consider the simple case of a scalar order parameter $\phi(\mathbf r, t)$ which describes the ordered state of a broken Ising or $Z_2$ symmetry. We note that the checkerboard CDW order of the square-lattice Holstein model, which will be discussed in Sec.~\ref{sec:esn-cdw}, is also described by a scalar order parameter. 

Phenomenologically, the time-evolution of the order parameter is often described by a partial differential equation (PDE). The specific form of the PDE depends on the symmetry and conservation laws of the order-parameter field.  For more complicated systems, such as the Holstein model to be discussed below, the dynamics of the emergent order parameter might not be explicitly described by a PDE. Moreover, while most PDEs can be numerically solved with linear-scaling method, the time complexity for phase-ordering simulations of complex electron systems is polynomial $\mathcal{O}(N^\alpha)$ with the exponent $\alpha$ greater than 1, where $N$ is the number of lattice points. For example, if the electron structure problem required to compute the driving force is solved using exact diagonalization, the computational time scales cubically with the system size.

The fact that the time-evolution is modeled by differential operators implies the dynamics is spatially local, which is also key to the linear-scaling method. Indeed, as argued by W. Kohn, linear-scaling electronic structure methods are possible mainly because of the locality nature or “nearsightedness” principle~\cite{kohn96,prodan05} of many-electron systems. For example, the locality principle was tacitly assumed in most ML interatomic potentials to achieve large-scale {\em ab initio} molecular dynamics simulations with desired quantum accuracy~\cite{behler07,bartok10,li15,botu17,smith17,zhang18,behler16,shapeev16,mcgibbon17,suwa19,chmiela17,chmiela18,sauceda20}. Similar ML force-field methods have also been recently developed to enable multi-scale dynamical modeling of several well-known condensed-matter lattice models~\cite{zhang20,zhang21,zhang22,zhang23,zhang22b,cheng23,cheng23b}.  


Based similarly on the locality principle, here we assume that the update to a local order parameter only depends on order-parameter configuration within an immediate neighborhood. Explicitly, this implies the following time evolution for a local order parameter $\phi_i(t) \equiv \phi(\mathbf r_i, t)$ at lattice site~$\mathbf r_i$:
\begin{eqnarray}
	\label{eq:phi-evolution1}
	\phi_i( t+ \Delta t) =  \mathcal{F}\bigl( \{ \phi_j( t) \, | \, \mathbf r_j \in \mathcal{N}_i \} \bigr).
\end{eqnarray}
Here $\mathcal{N}_i$ denotes a finite neighborhood block of the lattice centered at site-$i$. The linear size $\ell$ of the block is related to the locality of the dynamical process. The function $\mathcal{F}(\cdot)$ encodes the complex dependence of the next-step order parameter at site-$i$ on the current configuration within the neighborhood. Importantly, this complex dependence is to be learned by the ESN in our scalable ML framework. The computational details are illustrated in FIG.~\ref{fig:schematic}. First, the order parameters $\{ \phi_j(t) \} $ in the neighborhood $\mathcal{N}_j$ are flattened into an array $\mathbf x(t)$ which is the input to the ESN. Then the recurrent relation Eq.~(\ref{eq:dyn1}) is employed to obtain reservoir neurons at the next time step. The output neuron determined by linear transformation Eq.~(\ref{eq:dyn2}) then gives the central order-parameter $\phi_i(t + \Delta t)$ at the next time-step. Since the sizes of the input layer is fixed by the locality of the dynamical process and is independent of the lattice sizes to be simulated, the trained ESN can be applied to larger systems without rebuilding or retraining. The linear in time scalability results from that fact that a new configuration of the whole system is obtained by sequentially applying the same trained ESN to every site on the lattice.


We note that similar scalable ESN models have previously been developed to learn the spatiotemporal evolution of systems described by partial differential equations (PDEs). These include the phase ordering dynamics of an Ising order parameter field governed by either time-dependent Ginzburg-Landau (TDGL) equation or the Cahn-Hilliard-Cook equation~\cite{Chauhan23}, as well as the chaotic spatiotemporal dynamics of Kuramoto-Sivashinsky equation~\cite{Pathak18}. Compared with these previous works,  our approach highlights the generality of the framework and the importance of locality of the dynamical processes. Moreover, for applications to phase ordering dynamics, as will be discussed in the next section, we also emphasize the reproduction of overall statistical properties of phase ordering dynamics in addition to the short-term predictions. 


\section{ESN for Time-dependent Ginzburg-Landau equation}

\label{sec:TDGL}

As a warm-up and benchmark, we first apply the above ML framework to simulate the coarsening of a non-conserved scalar order parameter. Physically, this describes a system with broken $Z_2$ symmetry, as represented by the well-studied ferromagnetic Ising models $\mathcal{H} = -J \sum_{\langle ij \rangle} \sigma_i \sigma_j$, where $\sigma_i = \pm 1$ indicates the two possible spin directions at a lattice site-$i$, and $J > 0$ denotes the interaction strength between nearest-neighbor pairs $\langle ij \rangle$. The Ising model exhibits a time-reversal symmetry, namely the system energy is invariant under transformation $\sigma_i \to -\sigma_i$ of all spins. The breaking of this $Z_2$ symmetry at low temperatures leads to an ordered ferromagnetic state. The order parameter is essentially the magnetization of the system. A coarse-grained local order parameter for the ferromagnetic order is defined as $\phi(\mathbf r, t) = \langle \sigma_i \rangle$, where the notation $\langle \cdots \rangle$ indicates averaging Ising spins over a small block associated with $\mathbf r$.

For an Ising model quenched into the low temperature phase, the total magnetization is not conserved during the phase ordering. The dynamical evolution of the coarse-grained order parameter field $\phi(\mathbf r, t)$ is described by the time-dependent Ginzburg-Landau (TDGL) equation~\cite{Bray1994,Onuki2002,Puri2009}
\begin{eqnarray}
	\frac{\partial \phi}{\partial t} = \tau \phi - g \phi^3 + D \nabla^2 \phi, 
\end{eqnarray}
where $\tau, g$ and $D$ are positive coefficients depending on microscopic details of the model. The first two terms describe the tendency towards a nonzero local order parameter, while the third term in the form of a diffusion term favors uniformity of the ordered states.  Phenomenologically, TDGL also corresponds to the pure relaxational model-A dynamics $\partial \phi / \partial t = - \delta \mathcal{E} / \delta \phi$~\cite{Hohenberg77}, where $\mathcal{E}[\phi] = \int d\mathbf r \left[ \frac{D}{2} (\nabla \phi)^2 - \frac{\tau}{2} \phi^2 + \frac{g}{4} \phi^4 \right]$ is the effective Ginzburg-Landau energy functional of the order-parameter field. The energy functional is minimized by two uniform fields $\phi(\mathbf r) = \pm \phi_m$ related to each other by a $Z_2$ symmetry, where $\phi_m = \sqrt{\tau/ g}$. The TDGL can be explicitly derived from the mean-field treatment of the Ising models governed by either the Metropolis or Glauber dynamics~\cite{Puri2009}. But more generally, the TDGL can be applied to describe dynamics of emergent Ising order in systems with short-ranged interactions.

The cell dynamics methods~\cite{Oono88,puri88} provide a convenient and robust approach to model the time evolution of TDGL. Introducing a square lattice for the discretized 2D field and a small time step $\Delta t$ for the discretized time evolution, the update of a local order-parameter at lattice site-$i$ is given by
\begin{eqnarray}
	\label{eq:discretized-TDGL}
	\phi_i(t+ \Delta t) = D \left[ \langle\langle \phi_i(t) \rangle\rangle - \phi_i(t) \right] + f\left(\phi_i(t) \right),
\end{eqnarray}
where $\langle\langle \phi_i \rangle\rangle = \frac{1}{4} (\phi_{i+x} + \phi_{i-x} + \phi_{i+y} + \phi_{i-y})$ and $f(x)$ is an injection map that produces two local minima $\pm \phi_m$ of a single cell in the GL energy functional. In this work, we choose $f(x) = A \tanh(x)$. The first $D$ term corresponds to the discrete lattice Laplacian operator. 

\begin{figure}[t]
\centering
\includegraphics[width=0.95\columnwidth]{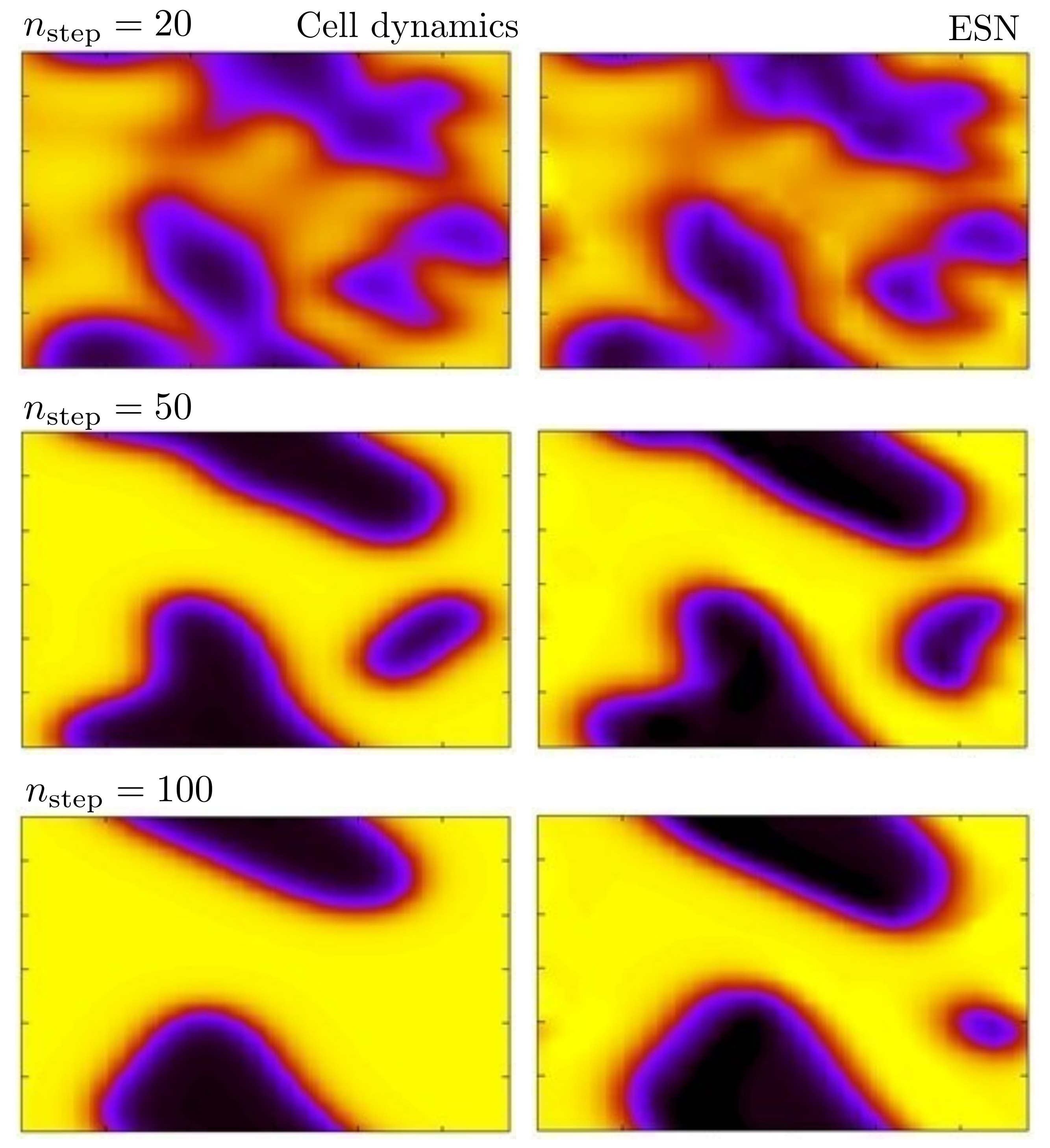}
\caption{Coarsening of Ising domains simulated by cell dynamics (left) and ESN (right). The color intensity shows the value of local order parameter $ \phi(\mathbf r, t) \in [-1, +1]$ at different times after a thermal quench. }
    \label{fig:ising-coarsening}
\end{figure}

In general, the discretized TDGL Eq.~(\ref{eq:discretized-TDGL}) indicates that updates of the local order parameter $\phi_i$ depend only on its first few neighbors depending on implementation details of the lattice Laplacian. In our case, this locality involves only the four nearest neighbors. As a result, we design an ESN model to predict $\phi_i(t + \Delta t) = y(t + \Delta t)$ with input only from the order parameters at the same site-$i$ and its four nearest neighbors, i.e. the input neurons are set to $\mathbf x(t) = \{ \phi_i(t), \phi_{i\pm x}(t), \phi_{i\pm y}(t) \}$. The reservoir layer consists of \(N=200\) nodes and is initialized with a sparse weight matrix $\mathbb{W}_{\rm res}$ of 60\% connectivity and a spectral radius $\rho = 0.79$. The elements of the input weight matrix \(\mathbb{W}_{\rm in}\) are randomly assigned with uniform distribution between~\(\pm0.25\).

The ESN was trained by a dataset from cell-dynamics simulations of a $30\times 30$ system with parameters $A = 1.3$ and $D = 0.5$. The dataset consists of time sequences of 100 steps from 30 randomly selected lattice sites along with the corresponding nearest neighbors. FIG.~\ref{fig:ising-coarsening} shows coarsening of Ising order generated by the cell dynamics method and ESN for time steps \(t=20\), \(t=50\), and \(t=100\) starting from the same initial random state. The ESN predicted coarsening agrees relatively well with that from the cell dynamics simulation for the early time steps, e.g. $t \le 50$. However, the ESN results start to deviate from the cell-dynamics simulations at late times. Similar results are reported in a recent study which shows rather accurate short-term predictions from ESN, and divergent trajectories at late times~\cite{Chauhan23}. Nonetheless, the overall domain geometries predicted by the ESN model, such as domain sizes and interface widths, remain similar to the actual behaviors. 

\begin{figure}
\centering
\includegraphics[width=0.99\columnwidth]{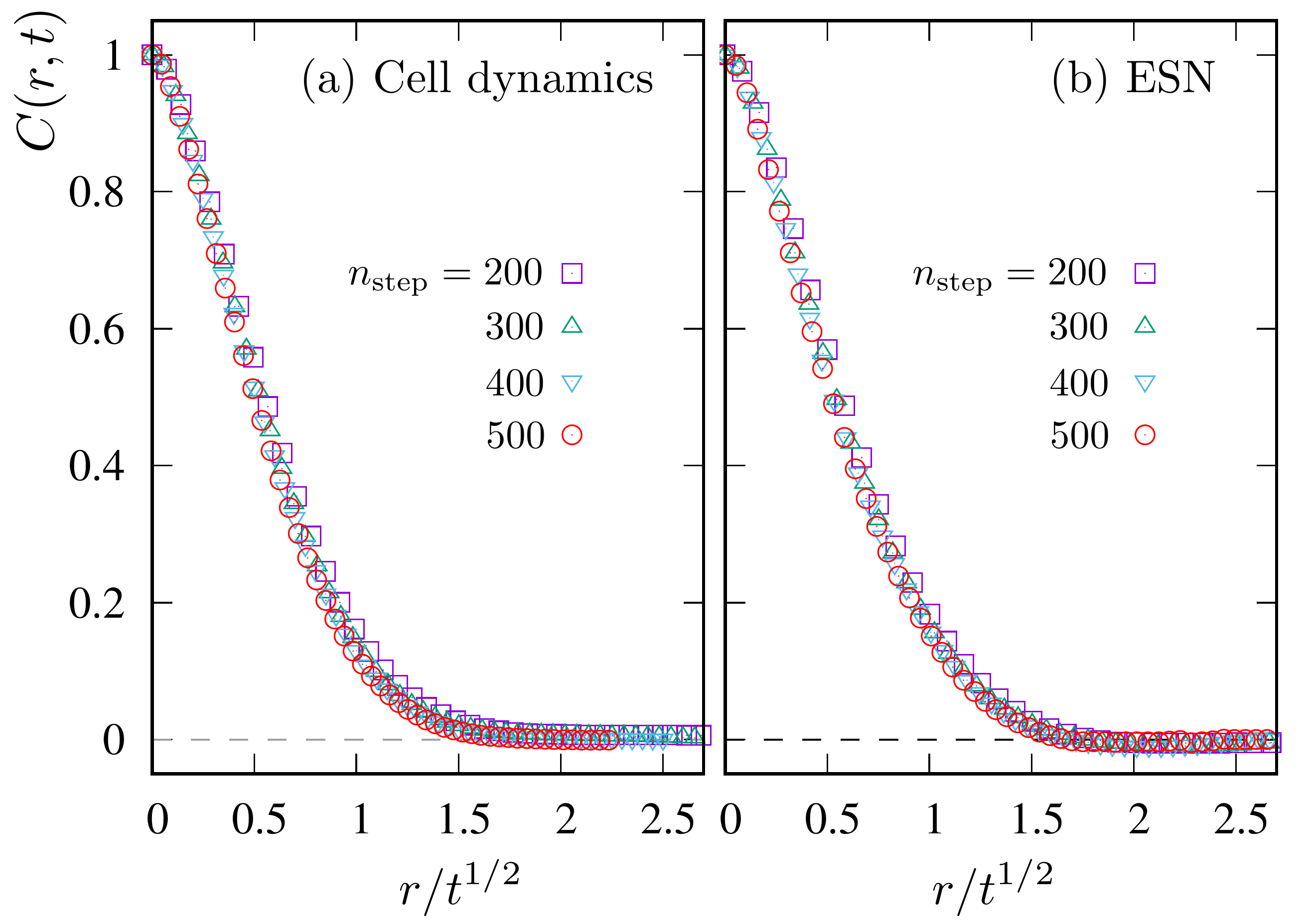}
\caption{Scaled correlation functions $C(r, t)$ versus $r / t^{1/2}$ based snapshots obtained from (a) cell dynamics simulations and (b) ESN predictions. }
    \label{fig:ising-corr}
\end{figure}

To quantitatively analyze the statistical properties of the coarsening dynamics, we performed cell dynamics simulations of the TDGL on a $100\times 100$ lattice with 50 different random initial conditions. The above ESN model trained from the $30\times 30$ dataset was also used to predict the time evolution of the same 50 initial states. The snapshots from both simulations were then used to compute the correlation function 
\begin{equation}
    C(\mathbf r ,t)=\frac{\langle \phi(\mathbf r_0, t)\phi(\mathbf r_0 + \mathbf r, t) \rangle -\langle \phi \rangle^2}{\langle \phi^2 \rangle -\langle \phi \rangle^2}.
    \label{correlation_func_cd}
\end{equation}
 FIG.~\ref{fig:ising-corr} shows the correlation functions at four different times from both cell-dynamics and ESN simulations. In both cases, by rescaling the distance by the square root of time, $r \to r / t^{1/2}$, the data points from different times nicely collapse on an underlying curve, indicating a dynamical scaling property of the domain coarsening~\cite{Bray1994,Onuki2002,Puri2009}. Specifically, this means that the correlation function depends on time through an emergent time-dependent length scale $L(t)$
\begin{eqnarray}
	\label{eq:dyn_scaling}
	C(\mathbf r, t) = F\left(\frac{\mathbf r }{ L(t)} \right)
\end{eqnarray}
where $F(x)$ is a universal scaling function, and the increase of this characteristic length, which can be viewed either as a correlation length or typical domain size, follows the Allen-Cahn growth law for non-conserved Ising order parameter~\cite{Allen1972}:	
\begin{eqnarray}
	L(t) \sim t^{1/2}.
\end{eqnarray}
Moreover, the scaling functions in the two cases also quantitatively agree with each other. This means that, even though the long-term predictions are not exact due to the statistical nature of ML approaches, the trained ESN successfully captures the statistical coarsening behavior of the TDGL.


\section{ESN for CDW dynamics of  semiclassical Holstein model}

\label{sec:esn-cdw}

In this section, we apply the ESN approach to predict the phase ordering dynamics of charge density wave (CDW) state in a square-lattice Holstein model. The checkerboard CDW order on a square lattice breaks the $Z_2$ sublattice symmetry and can also be described by a scalar Ising-type order parameter field. After performing proper benchmarks, our large-scale simulations enabled by the ESN models uncover an intriguing anomalous coarsening that is beyond the expected Allen-Cahn behavior. We first briefly review the CDW physics of the Holstein model.

\subsection{Holstein model: a brief review}

The Holstein model~\cite{Holstein1959} is a canonical system for studying electron-phonon physics such as CDW order, polaron dynamics, and phonon-induced superconductivity~\cite{bonca99,golez12,mishchenko14,scalettar89,costa18,bradley21}. It describes a lattice model of itinerant electrons interacting with scalar dynamical variables, which represent local $A_1$-type structural distortions associated with each lattice site. For bipartite lattices in both 2D and 3D, the half-filled electron band is unstable towards the formation of a checkerboard charge-density modulation, which is accompanied by a staggered arrangement of local lattice distortions~\cite{noack91,zhang19,chen19,hohenadler19}. The Hamiltonian of the Holstein model reads
\begin{align}
	& &  {\mathscr{H}} = -t_{\rm nn} \sum_{\langle ij \rangle}  {c}_i^\dagger  {c}_j - g \sum_i \left( c^\dagger_i c^{\,}_i - \frac{1}{2} \right) Q_i \notag\\
	& & \qquad  +\sum_i \left( \frac{P_i^2}{2 m} + \frac{k Q_i^2}{2} \right)+ \kappa \sum_{\langle ij \rangle} Q_i Q_j.
	\label{eq:H_holstein}
\end{align}
where $ {c}^\dagger_i$ ($ {c}_i$) is the electron creation (annihilation) operator at site-$i$, $Q_i$ represents a scalar dynamical lattice degree of freedom associated at the $i$-th lattice site, and $P_i$ is the corresponding momentum variable. The first term above describes electron hopping between a nearest-neighbor pairs $\langle ij \rangle$ on the lattice, with $t_{\rm nn}$ being the hopping coefficient. The second term describes the electron-lattice coupling, where $g$ is the coupling constant and $n_i = {c}^\dagger_i  {c}^{\,}_i$ is the electron number operator. The second line in the above Hamiltonian describes the classical energy of the lattice system: $m$ is the effective mass, $k$ is an on-site elastic constant, and $\kappa$ is the coupling coefficient between nearest-neighbor lattice variables. The Holstein model could also be used to describe real compounds, where $Q_i$ represent amplitudes of local collective modes of atomic clusters such as the breathing mode of a local oxygen octahedron enclosing a transition metal ion at site-$i$. 

\begin{figure}
\centering
\includegraphics[width=0.99\columnwidth]{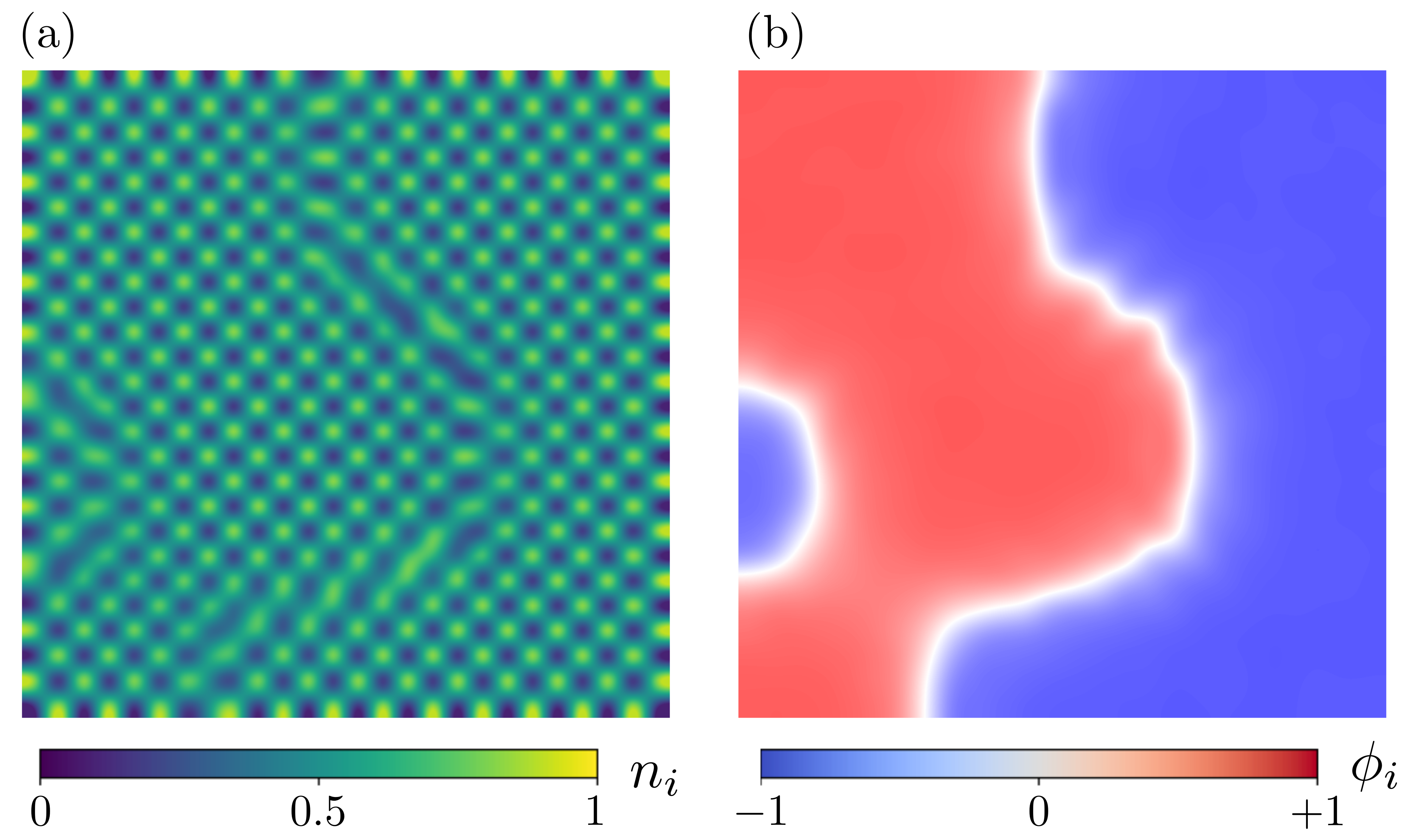}
\caption{(a) a snapshot of the on-site electron number $n_i = n(\mathbf r_i)$ obtained from ED-Langevin dynamics simulations. The corresponding local Ising-type CDW order parameter $\phi_i$ is shown in panel~(b).   }
    \label{fig:CDW-snapshot}
\end{figure}

The CDW order of half-filled Holstein model on bipartite lattices is characterized by a checkerboard electron density modulation; an example is shown in FIG.~\ref{fig:CDW-snapshot}(a).  Due to the electron-lattice coupling, the checkerboard charge modulation is accompanied by a staggered lattice distortion $Q_{A/B} = \pm \mathcal{Q}$. The CDW order of the Holstein model thus breaks the $Z_2$ sublattice symmetry of the bipartite square lattice.  
On symmetry ground, the CDW transition in the Holstein model is expected to belong to the Ising universality class, which is indeed confirmed by quantum Monte Carlo simulations~\cite{noack91,zhang19,chen19,hohenadler19}.  In order to characterize inhomogeneous CDW states during the phase ordering, we introduce a local Ising-like scalar order parameter  
\begin{align}
	\label{eq:phi-i}
	\phi_i = \Bigl(n_i - \frac{1}{4}\sum_j\phantom{}^{'} n_j \Bigr) \exp\left({i \mathbf Q \cdot \mathbf r_i}\right), 
\end{align}
where $\mathbf Q = (\pi, \pi)$ is the ordering wave vector of the checkerboard pattern, the phase factor $\exp({i \mathbf Q \cdot \mathbf r_i}) = \pm 1$ for the $A$ and $B$ sublattices, respectively, and the prime in the summation indicates that the sum is restricted to the nearest neighbors of site-$i$. This local order parameter essentially measures the difference of the electron number at a given site and that of its nearest neighbors. A nonzero $\phi_i$ thus indicates the presence of local charge modulation around site-$i$. FIG.~\ref{fig:CDW-snapshot}(b) shows the scalar order parameter corresponding to the CDW state in panel~(a). The red and blue regions, corresponding to $\phi_i =+1$ and $-1$, respectively, are CDW domains related by the $Z_2$ symmetry. The two types of CDW domains are separated by interfaces of vanishing $\phi_i$, corresponding to the white regions. 

While the transition to the CDW phase in Holstein model has been firmly established to belong to the Ising universality class, the kinetics of the CDW phase transition, however, is relatively less explored, even with the semiclassical approximation for the phonons. This is partly due to the computational complexity of dynamical simulations involving the itinerant electron degrees of freedom. Within the semiclassical approximation, the dynamics of the local lattice modes is governed by an effective Langevin equation
\begin{eqnarray}
	\label{eq:langevin}
	m\frac{d^2Q_i}{dt^2} = - \frac{\partial \langle  \mathscr{H} \rangle}{\partial Q_i}  - \gamma \frac{dQ_i}{dt} + \eta_i(t)
\end{eqnarray}
Here the Langevin thermostat is used to account for the effects of a thermal reservoir during the phase ordering; $\gamma$ is a damping constant and $\eta_i(t)$ is a thermal noise of zero mean and variance $\langle \eta_i(t) \eta_j(t') \rangle = 2 \gamma k_B T \delta_{ij} \delta(t - t')$.

While standard methods, such as the velocity-Verlet algorithm, can be straightforwardly used to integrate the equation of motion, the computational bottleneck is the calculation of the driving forces given in the first term on the right-hand side of Eq.~(\ref{eq:langevin}). In the adiabatic limit, the force calculations can be simplified by assuming a much faster electron relaxation time compared with the lattice dynamics, an approximation similar to the Born-Oppenheimer approximation in quantum molecular dynamics~\cite{Marx09}. However, this still requires solving the electron Hamiltonian in order to evaluate the driving forces coming from the electrons. For the semiclassical Holstein model which is quadratic in the electron operators, the electron Hamiltonian can be solved using exact diagonalization (ED) with a time-complexity of $\mathcal{O}(N^3)$. Large-scale dynamical simulations are thus computationally infeasible due to the expensive ED calculations required for every time-step. 

Efficient linear-scaling approaches to solving the electron structure problem can be achieved by two major methods. The first one is based on the kernel polynomial methods (KPMs) which offer an $\mathcal{O}(N)$ method to solve quadratic fermionic systems~\cite{Weisse06,Barros13,Wang18}. In this approach, relevant quantities are expanded in Chebyshev polynomial series, where the expansion coefficients can be efficiently computed using sparse-matrix-vector multiplications. As discussed in Sec.~\ref{sec:intro}, ML methods can also be used to realize linear-scaling methods for complex electronic systems based on the locality principle. In particular, a Behler-Parrinello type ML force-field model has been developed to enable large-scale adiabatic dynamics of CDW order in the Holstein model~\cite{cheng23}. In the following, we present another ML approach to this problem, focusing on the learning of dynamical time series through reservoir computing.

\subsection{ESN model for CDW coarsening}

It is worth noting that the CDW order parameter~$\phi_i$ defined in Eq.~(\ref{eq:phi-i}) represents an emergent entity of the Holstein system. The fundamental dynamical degrees of freedom in the adiabatic approximation is the on-site lattice distortion $Q_i$, which is governed by the Langevin dynamics. As a result, the time-evolution of CDW order implicitly follows the underlying adiabatic lattice dynamics. It is not even clear whether the effective dynamics for $\phi$ can be expressed in terms of a partial differential equation, in contrast to the case of TDGL for short-ranged Ising systems. The RNN, trained by time series of microscopic calculations, thus provides an effective empirical modeling of the emergent dynamics of the order-parameter field. 

Here we apply the scalable ESN-based scalable ML approach, depicted in FIG.~\ref{fig:schematic}, to learn the coarsening dynamics of the CDW order in semiclassical Holstein model. Again, based on the locality principle, the evolution of a local CDW order $\phi_i(t)$ is assumed to depend on the order-parameter configuration in its neighborhood as described in Eq.~(\ref{eq:phi-evolution1}). As discussed above, the effective interaction between the lattice distortions $Q_i$ is mediated by the itinerant electrons and, as a result, is longer ranged. This in turn implies the need for a greater neighborhood area $\mathcal{N}_i$ on which the local evolution depends for the ESN implementation, in contrast to the case of TDGL discussed in Sec.~\ref{sec:TDGL} with only the four nearest neighbors in addition to the center site in the input.

To account for the larger range of locality, a $7\times 7$ block centered at site-$i$ is designated as the neighborhood~$\mathcal{N}_i$ in our ESN model for the CDW dynamics. The size of the reservoir network is set at $N = 200$ neurons. And as before, the single neuron at the output is the predicted local CDW order parameter $\phi_i(t + \Delta t)$ of the next time step for the selected lattice site. The reservoir weight matrix $\mathbb{W}_{\rm res}$ was generated with a sparsity of 15\% connectivity and the nonzero values were initialized from sampling a normal distribution of zero mean and unit variance.  A rescaling procedure discussed above was then carried out to produce a spectral radius of $\rho = 0.79$. 

In order to properly incorporate the lattice symmetry into the ESN model, special care was taken for the initialization of the input weight matrix $\mathbb{W}_{\rm in}$. The symmetry for a given lattice site-$i$ is described by the $D_4$ point group in the case of square lattice.  This means that two neighborhood configurations $\{ \phi^{(1)}_j(t) \}$ and $\{ \phi^{(2)}_j(t) \}$ that are related by symmetry operations of $D_4$ should produce exactly the same output $\phi_i(t + \Delta t)$ at the center site. To this end, we first note that  neighboring sites with the same distance from the center are mapped to each other under the symmetry operations. The input weights are constrained by the following form
\begin{eqnarray}
	\label{eq:W_in_sym}
	(\mathbb{W}_{\rm in})_{\alpha j} = w_{\alpha,  r_{ij} } \, f(r_{ij}),
\end{eqnarray}
where $r_{ij} = |\mathbf r_j - \mathbf r_i|$ is the distance from the center site-$i$, $\alpha = 1, 2, \cdots, N_{\rm res}$ is an index for neurons in the reservoir, $w_{\alpha, r}$ represent a set of random numbers depending on reservoir neuron index $\alpha$ and a distance $r$, and $f(r_{ij})$ denotes an influence function. It should be noted that, contrary to independent random initialization of each matrix element of $\mathbb{W}_{\rm in}$, all neighbors with the same distance $r_{ij}$ to the center now share the same ``random" weight $w_{\alpha, \, r_{ij} }$ to the $\alpha$-th neuron in the reservoir. In our implementations, these independent random numbers $w_{\alpha, r}$ were similarly initialized from a normal distribution of zero mean and unit variance. We have also explicitly demonstrated the importance of this input symmetrization by showing that the prediction accuracy of ESNs with input weight constrained by Eq.~(\ref{eq:W_in_sym}) is significantly better than that of ESN with a unconstrained random $\mathbb{W}_{\rm in}$. 

Finally, the function $f(r)$ is introduced to account for the different influences of neighboring sites. It is expected that closer neighbors would have larger influence than furthest ones.  A simple decay function $f(r) = 1/r$ was used in our implementation.

The dataset for training the ESN was generated from one single ED-Langevin dynamics simulation of the Holstein model on a $40\times 40$ lattice. The filling fraction is set at $f = 0.5$, i.e. one electron per lattice sites. We consider a thermal quench scenario where an initial state with random distortions $Q_i$ is suddenly quenched to a low temperature $T = 0.001$ at time $t = 0$. 
By setting the nearest-neighbor hopping coefficient $t_{\rm nn} = 1$, which serves as the energy unit, the other parameters of the Holstein model are $g = 1$, $k=1$, and $\kappa=0.18$. The lattice dynamics is characterized by the fundamental frequency $\omega = \sqrt{k / m}$ of the simple harmonic oscillator. The fundamental oscillation period $t_0 = 2\pi / \omega$ will be used as the time unit in the following. The dissipation timescale is given by $\tau = \gamma / k$, and we have used a damping coefficient $\gamma$ such that the dimensionless dissipation in one fundamental cycle is $\omega \tau \sim 0.05$. Finally, an integration time-step $\delta t = 0.0025\, t_0$ was used in all Langevin dynamics simulations discussed in this work.

As the global error of the Verlet algorithm scales as~$\mathcal{O}(\delta t^2)$, a small time-step $\delta t$ was used in order to reduce the discretization errors in the integration of the ED-Langevin dynamics. The ESN model, on the other hand, is not subject to such discretization error and can be designed to predict time evolution at a relative larger time step. Yet, the prediction time-step $\Delta t$, which is constrained by the temporal resolution of the input/output time series, needs to be small enough in order to capture more detailed temporal features. As a trade-off, here we chose an ESN time-step that corresponds to 20 integration time step, i.e. $\Delta t = 20\, \delta t = 0.05 \, t_0$.


For the training dataset, a total of 400 CDW configurations separated by a time-step $\Delta t$ were collected from the ED-Langevin simulations of a $40\times 40$ lattice. For a given selected site-$i$,  its 400 CDW order parameter $\phi_i(m \Delta t)$ and the corresponding neighbor configurations $\{ \phi_j(m \Delta t) \, | \, j \in \mathcal{N}_i \}$, where $m = 1, 2, \cdots, 400$, are concatenated into a training sequence. This is repeated for 100 randomly selected lattice sites. To account for the input containing multiple time series of different lattice sites, before resuming training on the next lattice site's time series, we reinitialize the reservoir neurons and allow them to washout again.  Training is done using Eq.~(\ref{eq:dyn1}) as the state update equation with $f_{\rm av}(x)= \tanh(x)$ as the activation function. The predicted output time sequences are computed using Eq.~(\ref{eq:dyn2}), from which the Ridge regression method is employed to compute the trainable output weight $\mathbb{W}_{\rm out}$ according to Eq.~(\ref{eq:solve_Wout}). 


\begin{figure}
\centering
\includegraphics[width=0.9\columnwidth]{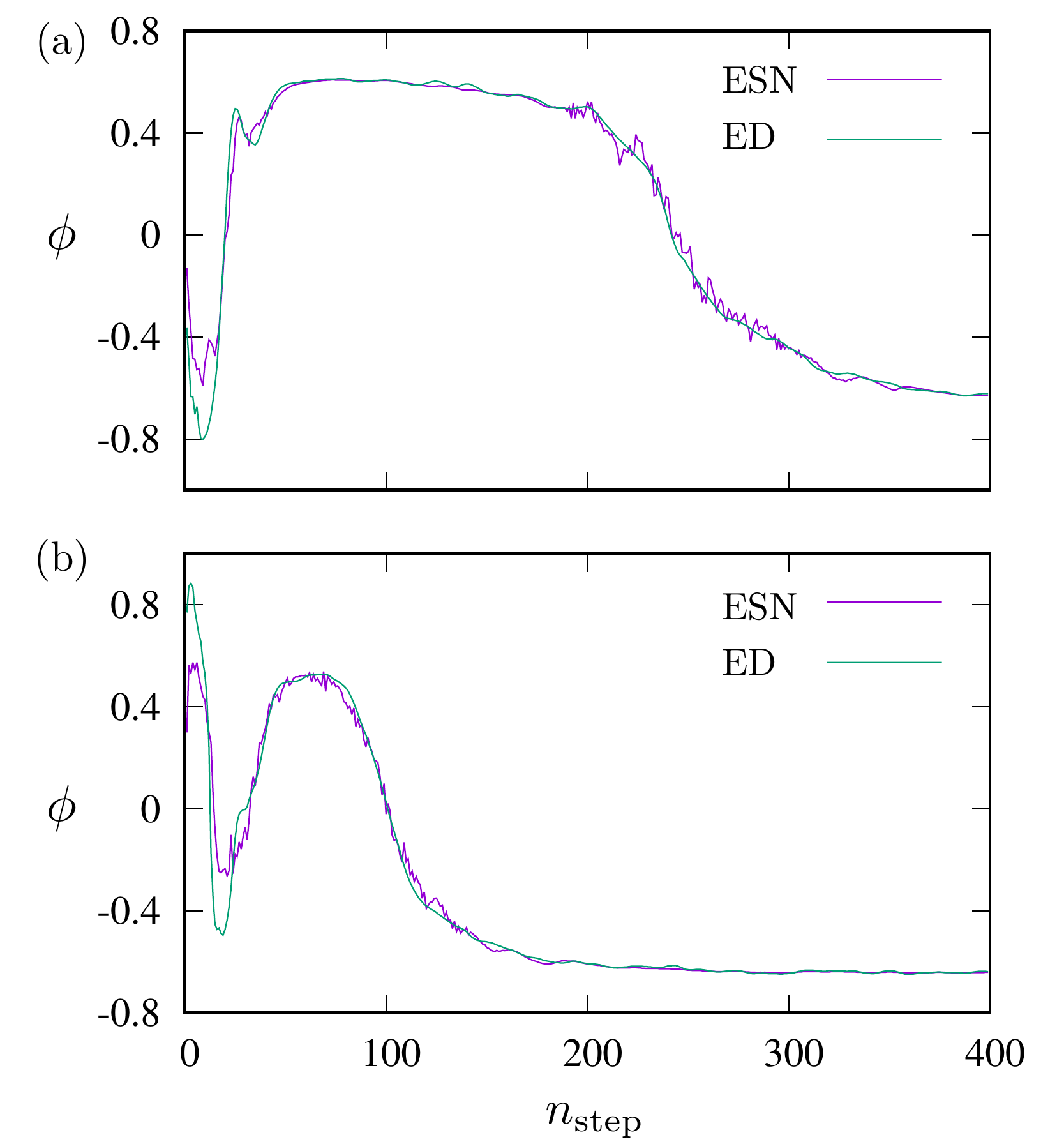}
\caption{Panels $(a)$ and $(b)$ represent the benchmark of ESN predictions for time series of local CDW order parameter $\phi_i(t)$ at two randomly selected sites.}
    \label{fig:benchmark1}
\end{figure}

As a first benchmark, we compare the time series of the local order parameter $\phi_i$ predicted by ESN against that obtained from ED-Langevin dynamics simulations on a relatively small $40\times 40$ lattice. The comparison of the local CDW order at the 400 time steps separated by $\Delta t$ are shown in FIG.~\ref{fig:benchmark1} for two randomly selected lattice sites. For this single-site dynamics benchmark, the CDW order $\phi_i(t)$ at the selected site-$i$ is recurrently fed into the ESN model for predicting the value at the next time-step $\phi_i(t + \Delta t)$, while the CDW configurations $\{ \phi_j(t) \}$ in the neighborhood, which are required input to the ESN model, are fixed by the exact time sequences from ED-Langevin simulations. The single-site time series is characterized by periods of roughly constant order parameter sandwiched by intermittent fast swings. A segment with a roughly constant $\phi_i$ indicates that site-$i$ is inside a well-developed ordered domain for the time period. The fast transition from positive to negative values, and vice versa, corresponds to a domain-wall passing through the site.

Our results show that the ESN can predict the time series to high accuracy when the neighborhood configurations are clamped by the exact data. In particular, the fast transitions due to a propagating domain-wall is well captured by the ESN predictions.  The nice agreements with the actual data mean that the exact time sequences of the neighborhood excite the proper dynamics of the reservoir, which in turn produce the accurate time series of the CDW order at the center site. It is also worth noting that for this constrained benchmark, namely with the neighborhood clamped at the actual data, the trained ESN is capable of accurately predicting not only the short-term but also the long-term behaviors. 

The next step in our ML framework, shown in FIG.~\ref{fig:schematic}, is to predict the phase ordering behaviors of a whole lattice based on the trained ESN for local dynamics. That means, contrary to the constrained benchmark discussed above, the configurations $\{\phi_j(t) \}$ in the neighborhood $\mathcal{N}_i$ that are input to the ESN for the prediction of $\phi_i(t + \Delta t)$ are themselves ESN predictions. This leads to a faster accumulation of prediction errors, which in turn results in a divergence of long-term predictions from actual time series. Indeed, the time evolution of the whole order-parameter field $\phi(\mathbf r_i, t)$ predicted by the trained ESN model quickly deviates from that of ED-Langevin simulations on the $40\times 40$ lattice. Compared with the case of TDGL shown in FIG.~\ref{fig:ising-coarsening}, significant deviations occur at a much earlier time-scale in the ESN predictions of the CDW coarsening.

\begin{figure}
\centering
\includegraphics[width=0.99\columnwidth]{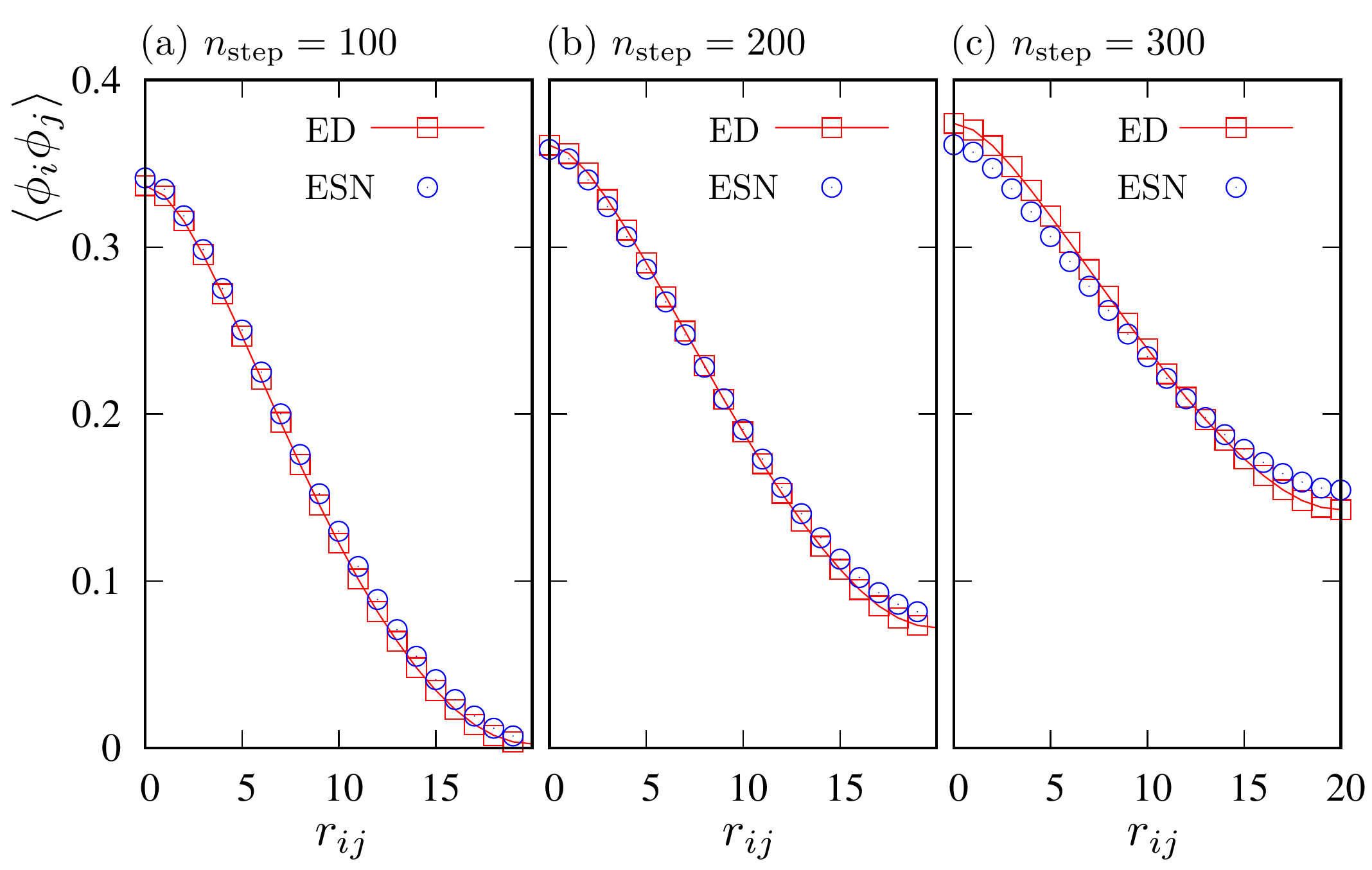}
\caption{Panels $(a)\mathbf{-}(c)$ are order-parameter correlations $\langle \phi_i(t) \phi_j(t) \rangle$ at three different times after a thermal quench. The curves were computed from ED-Langevin simulations and ESN predictions on a $40\times 40$ lattice.}
    \label{fig:benchmark-corr}
\end{figure}

Nonetheless, as in the case of TDGL, the ESN-based ML approach still captures the statistical behaviors of the CDW coarsening. This is demonstrated in FIG.~\ref{fig:benchmark-corr} which shows the bare correlation functions $\langle \phi_i(t) \phi_j(t) \rangle$ obtained from ESN and ED-Langevin simulations on a $40\times 40$ lattice. For both methods, 30 quench simulations with different initial conditions were performed to generate order-parameter fields for the calculation of the correlations. Although the bare correlation decays to zero with large separation at earlier times after the quench, system-wide ordered domains quickly develop because of the relatively small lattice sizes, giving rise to nonzero correlation even at $r_{ij} = L/2$ at late times; see FIG.~\ref{fig:benchmark-corr}(b) and (c). Importantly, the correlation functions from ESN simulations agree very well with those from exact simulations for all three times after the quench.

\subsection{Anomalous coarsening of CDW domains}

\begin{figure*}
\centering
\includegraphics[width=1.99\columnwidth]{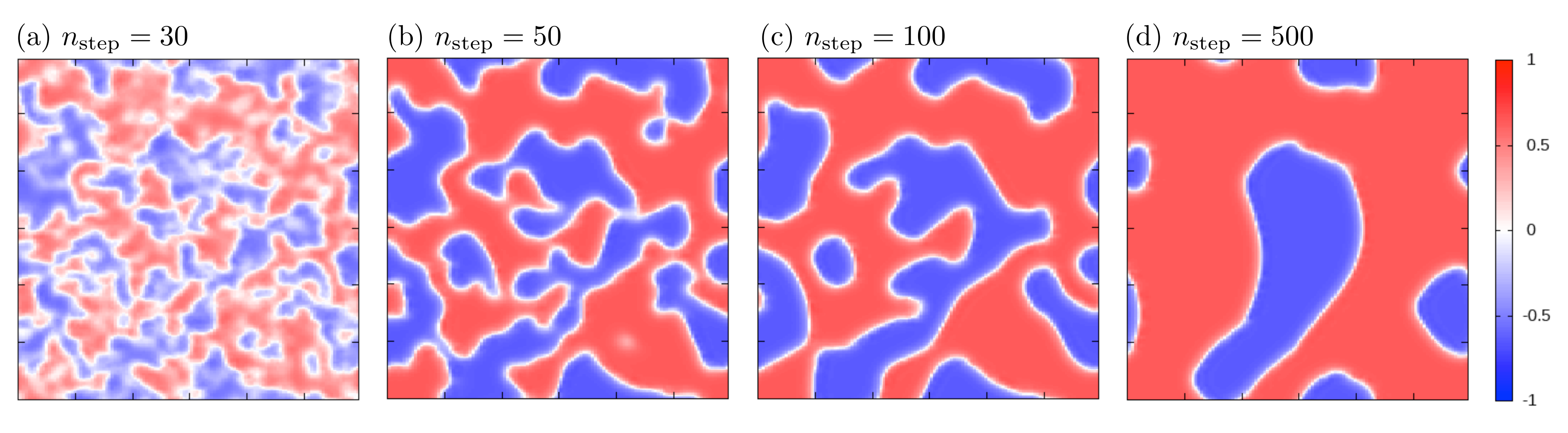}
\caption{Panels $(a)\mathbf{-}(d) $ represent ESN simulation of the coarsening of CDW domains on a $120 \times 120$ lattice. The ESN is trained by dataset generated from one ED-based quench simulation on a $40\times 40$ lattice. The ESN model was designed to capture the phase ordering dynamics of the Holstein model at a low temperature $T = 0.001$.  }
    \label{fig:coarsening-cdw}
\end{figure*}

Next we apply benchmarked ESN model to study large-scale coarsening dynamics of CDW domains of the Holstein model. As discussed above, large-scale dynamical simulations are possible mainly because of the efficient linear scalability of ML methods. As a demonstration, ED-Langevin dynamics simulation of a $40\times 40$ Holstein model took about 16 hours for $10,000$  integration steps~($\delta t$). On the other hand, ESN-based simulations of the same system size and $500$ prediction steps~($\Delta t$), corresponding to the 10,000 $\delta t$ integration steps, only took only about 5 seconds, meaning a roughly $10^4$-fold improvement in efficiency.  Extrapolating to simulations on a $120\times 120$ lattice to be discussed below based on the $\mathcal{O}(N^3)$ scaling, the ML-based simulations are expected to be roughly $10^7$ times faster than that based on the ED method. Even compared with the Langevin simulations based on the linear-scaling KPM discussed above, the ESN model is still much more efficient thanks to faster computations of the reservoir network and the advantage of a larger prediction step.

\begin{figure} [b]
\centering
\includegraphics[width=0.99\columnwidth]{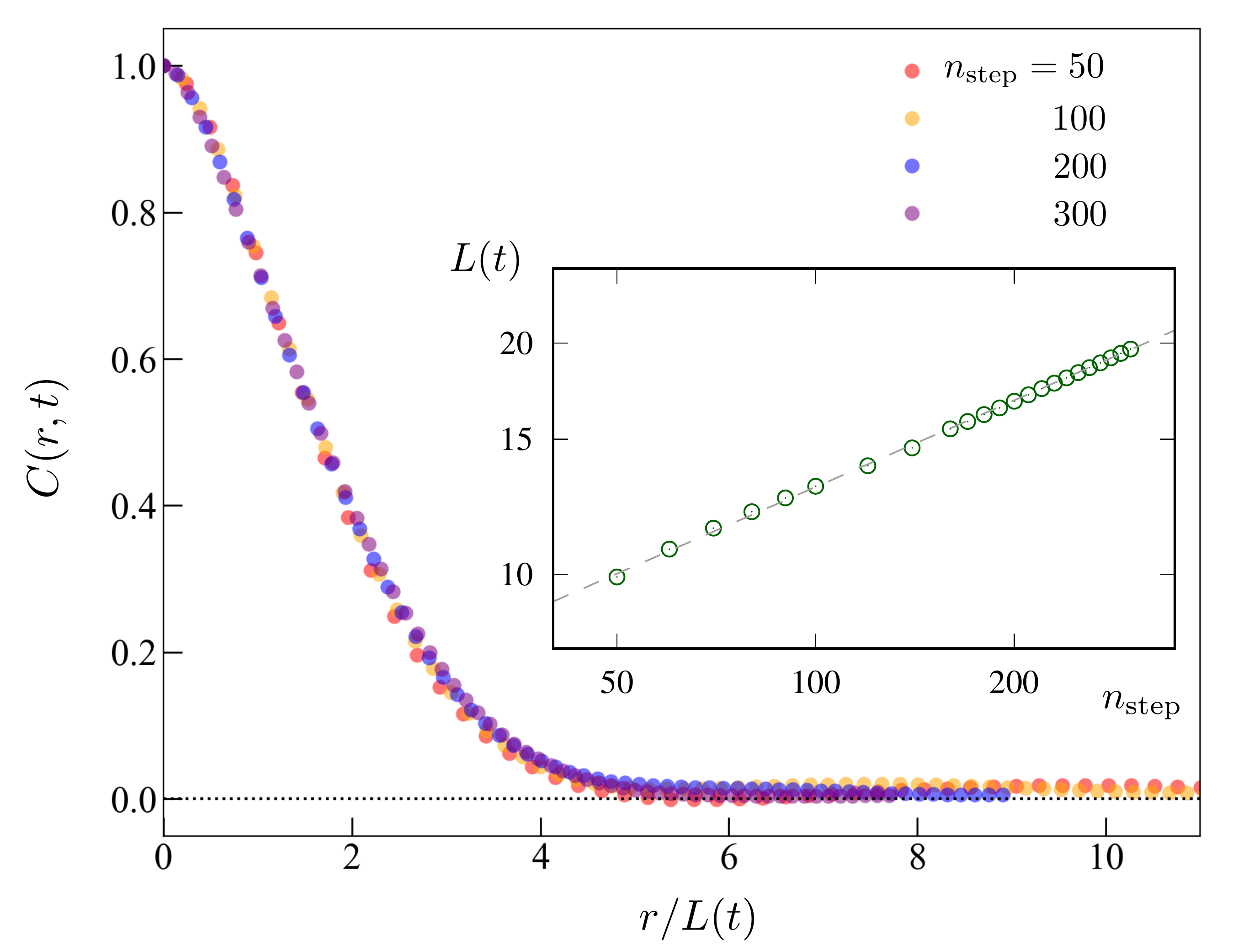}
\caption{Correlation function of the CDW order parameter at different times versus a distance $r / L(t)$ rescaled by the time-dependent correlation length. The distance is measured along the $x$ axis, with averaging also from data along the $y$-axis. The inset shows the correlation length $L(t)$ versus time in a log-log plot. The simulation time is related to number of steps as $t = n_{\rm step} \Delta t$.  }
    \label{fig:corr-scaling}
\end{figure}

FIG.~\ref{fig:coarsening-cdw} shows snapshots of the CDW order-parameter field $\phi_i$ predicted by ESN at four different times after a quench. The red and blue regions, corresponding to $\phi_i =+1$ and $-1$, respectively, are CDW domains related by the $Z_2$ symmetry. The two types of CDW orders are separated by interfaces of vanishing $\phi_i$, corresponding to the white regions. The early-stage relaxation of the system is characterized by the development of local CDW order, as indicated by the increasing amplitude $|\phi_i|$ of the order parameter. The relative small sizes of CDW domains, however, indicates a rather short correlation length.  As the system relaxes toward equilibrium, these CDW domains merge into larger ones, giving rise to a coarser mixture of the two ordered phases.

To quantify the dynamics of domain growth, we first compute the properly normalized correlation function of the CDW order parameter according to Eq.~(\ref{correlation_func_cd}). A time-dependent characteristic domain size can be estimated from the correlation length defined as
\begin{eqnarray}
	L(t) = \frac{ \sum_{\mathbf r} | \mathbf r | C(\mathbf r, t) }{ \sum_{\mathbf r} C(\mathbf r, t) }.
\end{eqnarray}	
For example, for an exponentially decaying correlation function $C(r) \sim \exp(-r/\ell)$, the correlation length computed using the above formula gives the exact decay length $L = \ell$. FIG.~\ref{fig:corr-scaling} shows the CDW correlation functions at different time steps versus a distance $r/ L(t)$ rescaled by the computed $L(t)$. The nearly perfect data-point collapsing indicates the coarsening of CDW domains exhibit a dynamical scaling symmetry~\cite{Bray1994,Onuki2002,Puri2009}. This means that the time-dependent correlation function can be expressed in the form of Eq.~(\ref{eq:dyn_scaling}), with the scaling function $F(x)$ given by the collapsed data points.

The presence of a dynamical scaling also indicates that the CDW domain configurations exhibit similar structures at different times and, as a result, the coarsening dynamics is encoded in the time dependence of the correlation length $L(t)$. By plotting the numerically computed $L(t)$ as a function of time in a log-log plot, as shown in the inset of FIG.~\ref{fig:corr-scaling}, we find a power-law domain growth
\begin{eqnarray}
	L(t) \sim t^{\alpha},
\end{eqnarray}
with an exponent $\alpha = 0.375$ that is smaller than the Allen-Cahn value $\alpha = 1/2$ expected for a non-conserved Ising order parameter. The smaller growth exponent, however, is consistent with results from previous works based on the ML force-field approach~\cite{cheng23}. To understand this unusual coarsening behavior, we note that the Allen-Cahn square-root law is physically based on the assumption that the domain-wall velocity is proportional to the corresponding local curvature~\cite{Allen1972}. This result can also be obtained from the TDGL theory, which describes short-range interaction models with Ising symmetry. On the other hand, as discussed above, the effective interactions in the Holstein model arise from the mediation of itinerant electrons. Consequently, the Allen-Cahn theory likely cannot be applied to the coarsening of CDW domains as the resultant effective interactions are longer-ranged and frustrated.

\section{Conclusion and Outlook}

\label{sec:conclusion}

To summarize, we have developed a scalable ML framework based on Echo State Network (ESN) for model-free prediction of adiabatic phase ordering dynamics in complex electron systems. The approach is demonstrated on the coarsening dynamics of charge-density wave (CDW) orders in the square-lattice Holstein model,  where a checkerboard electron density modulation at half-filling is stabilized by a commensurate lattice distortion. The ESN uses local CDW order-parameters from a neighborhood around a given site as input, and predicts the CDW order at the center site for the next time step. Special attention is given to designing the connections between the hidden layer and output node to preserve lattice symmetries in the model. The linear scalability of our ML-ESN approach is fundamentally based on the locality principle, which states that the update of a local order parameter depends solely on its immediate surroundings. Because the predictions depend solely on CDW configurations within a finite region, the ESN is both scalable and transferable, enabling a model trained on smaller systems to be applied directly to larger lattices.

We note that similar scalable ESN models have previously been developed to learn the spatiotemporal evolution of systems described by partial differential equations (PDEs). These include the phase ordering dynamics of an Ising order parameter field governed by either time-dependent Ginzburg-Landau (TDGL) equation or the Cahn-Hilliard-Cook equation, as well as the chaotic spatiotemporal dynamics of Kuramoto-Sivashinsky equation. The locality is implied by the fact that the dynamics of these systems is described by a PDE. In contrast to these previous works, there is no explicit differential equations governing the dynamics of CDW order in the Holstein model. As an emergent entity, the time-evolution of CDW order implicitly follows the underlying adiabatic lattice dynamics. Our work thus provides an important proof-of-concept that ESN is capable of learning and predicting {\em emergent} dynamics of order-parameter field in complex electron systems. In particular, our demonstration opens the promising potential of utilizing ESN to learn coarsening dynamics from experimental data.

The ESN-based ML framework presented in this work is also similar to the popular ML force-field approaches. In both cases, the linear-scalability relies on the locality principle of physical systems to be modeled. The important difference is that the time evolution in the ML force-field approach is based on physical dynamics, be it the Langevin equation in MD or Landau-Lifshitz-Gilbert equation for spin dynamics. The focus of the ML model is on learning the time-consuming force calculations that drive the dynamics, while conventional numerical techniques are used to integrate the equation of motion with the ML-predicted forces. ML force-field models thus represent a more constrained ML approach to dynamical modeling. However, the ESN model is not designed to merely learn the combined effects of these two steps (force calculation and one-step $\delta t$ integration).  On the other hand, since the prediction step $\Delta t$ of ESN could be much larger than the integration time-step as discussed above, the ESN is trained to capture the time sequence of local order-parameters, regardless of whether the underlying dynamics is governed by a differential equation.

\bigskip

\begin{acknowledgments}
The authors thank Sheng Zhang for useful discussions. The work was supported by the US Department of Energy Basic Energy Sciences under Contract No.~DE-SC0020330. The authors also acknowledge the support of Research Computing at the University of Virginia.
\end{acknowledgments}

\bibliography{ref}

\end{document}